\documentclass[aps,prl,reprint,onecolumn]{article}
\usepackage[margin=1.in,nomarginpar]{geometry}
\usepackage{url}
\usepackage{graphicx}
\usepackage{pdflscape}
\graphicspath{ {figures/} }
\usepackage{caption}
\usepackage{subcaption}
\usepackage[nolist, nohyperlinks]{acronym}
\begin{acronym}[TDMA]
\acro{AI}{artificial intelligence}
\acro{FP}{Fabry-Perot}
\acro{PDH}{Pound-Drever-Hall}
\acro{DoF}{degree of freedom}
\acro{RL}{reinforcement learning}
\acro{DRL}{deep reinforcement learning}
\acro{Meta-RL}{meta reinforcement learning}
\acro{ML}{machine learning}
\acro{MDP}{Markov decision process}
\acro{GW}{gravitational wave}
\acro{RT}{real-time}
\acro{DDPG}{deep deterministic policy gradient}
\acro{TD3}{Twin Delayed DDPG}
\acro{SAC}{Soft Actor Critic}
\acro{PPO}{Proximal Policy Optimization}
\acro{RNN}{recurrent neural network}
\acro{GRU}{gated recurrent unit}
\acro{DPG}{deterministic policy gradient}
\acro{DQN}{deep Q-network}
\acro{TDS}{time delay system}
\acro{ADC}{analogue-to-digital converter}
\acro{DAC}{digital-to-analogue converter}
\acro{DSP}{digital signal processor}
\acro{EuCAIF}{European Coalition for AI in Fundamental Physics}
\acro{EuCAIFCon}{EuCAIFCon 2025}
\acro{FPGA}{field programmable gate array}
\acro{JIT}{just-in-time}
\acro{FSR}{Free Spectral Range}
\end{acronym}
\usepackage{amsmath}
\usepackage[scientific-notation = false]{siunitx}
\DeclareSIUnit\flops{FLOPS}
\usepackage{hyperref} 
\usepackage{algorithm}
\usepackage{algpseudocode}
\usepackage{longtable}
\usepackage{makecell}  
\usepackage{authblk}
\usepackage[
  backend=biber,
  style=numeric-comp,
  sorting=none,
  doi=true,
  isbn=false,
  eprint=true,
  url=false,
  giveninits=true,
  maxnames=3,
  minnames=1
]{biblatex}
\hypersetup{pdfinfo={
Title={Title},
Author={Mateusz Bawaj},
Author={Andrea Svizzeretto}}
}
\begin{document}

\title{Adaptive time-domain simulation of optical cavities with arbitrary dynamics}

\author[1]{A.~Svizzeretto}
\author[2]{J.~Casanueva~Diaz}
\author[3]{B.~L.~Swinkels}
\author[1]{M.~Bawaj\thanks{mateusz.bawaj@unipg.it}}

\affil[1]{Universit\`{a} di Perugia, I-06123 Perugia, Italy}
\affil[2]{European Gravitational Observatory (EGO), I-56021 Cascina, Pisa, Italy}
\affil[3]{Nikhef, 1098~XG Amsterdam, Netherlands}

\date{\today}

\maketitle

\begin{abstract}
We present a fast time-domain simulator for optical cavities capable of reproducing non-linear dynamical regimes arising from ring-down effect during resonance crossings at high mirror velocities. The model is based on a recursive formulation of the intracavity electric field as a sum over round trips, preserving the cavity memory while maintaining high computational efficiency. The simulator is designed to achieve three main goals. First, the boundary conditions of the cavity can be modified at each simulation step, allowing arbitrary time-dependent variations of both mirror positions and input electric field. Second, the sampling frequency can be flexibly chosen by the user, however, it is internally adjusted before effectively executing the simulation to remain consistent with the cavity round-trip structure. Finally, high computational efficiency was obtained by avoiding the repeated evaluation of the full electric field history. The framework is validated through comparison with experimental data from the Virgo interferometer during a mechanical excitation experiment, showing good agreement in non-adiabatic regimes. Due to its efficiency and flexibility, the simulator provides a versatile tool for time-domain studies of optical resonators and future applications in real-time control and reinforcement-learning-based lock acquisition.
\end{abstract}

\section{Introduction}
\label{sec:Introduction}
Time-domain simulation of optical cavities is required whenever the assumptions underlying frequency-domain or adiabatic models break down, such as in the presence of fast transients, short optical pulses, or rapidly varying boundary conditions. In these regimes, the intra-cavity field evolution depends explicitly on the finite round-trip time and the storage of optical energy~\cite{Lawrence_1999}.

This is particularly relevant in precision interferometry and gravitational-wave detectors, where the interaction between delayed intra-cavity fields and moving mirrors leads to non-linear, history-dependent dynamics that cannot be captured by stationary transfer functions~\cite{Bhawal_1998, Beausoleil_1999}.

Beyond interferometric gravitational-wave detectors, similar requirements arise in pulsed cavity ring-down spectroscopy, fast cavity tuning experiments, and active stabilization of high-finesse resonators, where transient dynamics and non-adiabatic effects determine system performance~\cite{Yong_Lee_1999}. In all these contexts, such simulations provide direct access to transient regimes and provide a natural interface for modern data-driven control strategies.

Legacy interferometer simulation frameworks like \texttt{e2e} and \texttt{SIESTA} provide detailed optical modelling capabilities~\cite{e2e_Bhawal_2000,siesta_Caron_1999}. However, their use in real-time and reinforcement-learning applications is limited by their complexity and software design. The simulator presented here is designed to address transient regimes regimes with minimal approximations for a wide range of cavity parameters. It explicitly propagates the intra-cavity field as a sum over round trips, retaining the full temporal memory of the cavity~\cite{Rakhmanov2000,svizzeretto}. Crucially, both the input optical field and the mirror positions can be modified at every simulation step. This allows arbitrary laser amplitude and phase modulation, as long as their characteristic frequencies remain below the simulator sampling frequency, as well as time-dependent cavity length variations under the same condition.

This level of flexibility is essential for our primary target application: integration into a Gymnasium-compatible reinforcement-learning environment for training real-time \ac{AI} agents dedicated to lock acquisition of suspended optical cavities currently under development~\cite{Svizzeretto_2026,bawaj2026_arxiv}. Lock acquisition is inherently a non-stationary control problem, dominated by transients, delays, and non-linear field-mirror coupling~\cite{Hassen_2014}. Its stepwise formulation naturally interfaces with discrete-time control algorithms, enabling realistic training conditions and supporting an effective sim-to-real transfer~\cite{Zhao}.

In addition to the theoretical and numerical developments presented here, the simulator is released as open-source software~\cite{oreonspy} under a separate license to ensure transparency and reproducibility.\\

The remainder of the paper is organized as follows.
Sec.~\ref{sec:Algorithm} introduces the numerical framework and the recursive formulation used to propagate the intracavity field, together with the implementation details of the simulation strategy. In particular we discuss the adaptive selection of the sampling frequency and the different integration regimes.

Sec.~\ref{sec:Validation_and_benchmarks} presents the validation of the simulator through a comparison with ring down experimental data from the arm cavity of the Virgo interferometer cavity, while Sec.~\ref{sec:Optical and Photonic relevance} extends the discussion to the broader applicability of the time-domain formulation to optics and photonics in general, as well as to \ac{ML}-aided control strategies.

Finally, Sec.~\ref{sec:Conclusions_and_Applications} summarizes the main results and outlines potential applications of the simulator, particularly in the context of real-time control and reinforcement-learning-based lock acquisition strategies.

Additional supporting material is provided in the Appendix, including the full comparison between simulated and experimental data from Virgo auxiliary channel over a \SI{10}{\second} time window, a reproduction of the pulsed cavity ring-down regimes reported by~\cite{Yong_Lee_1999}, and the complete benchmark table used to characterize the computational performance of the simulator.

\section{Algorithm}
\label{sec:Algorithm}

The algorithm of the simulator is capable of reproducing non-linear dynamical effects in optical cavities caused by the ring-down phenomenon. The temporal evolution of the electric cavity field is described in~\cite{Rakhmanov2000} by the recursive equation:
\begin{equation}
\begin{split}
    E(t) = t_{a}\sum_{n=0}^{N-1}\left[(r_{a}r_{b})^{n}\,e^{-2i k S_{n}(t)}\,E_{\mathrm{in}}(t-2n T)\right]+(r_{a}r_{b})^{N}\,e^{-2i k S_{N}(t)}\,E(t-2NT) \,,
    \label{eq:1.51}
\end{split}
\end{equation}
where, $t_a$ is the amplitude transmissivity of the input mirror, $r_a$ and $r_b$ are the amplitude reflectivities of the input and end mirrors, respectively, $k=2\pi/\lambda$ is the optical wave-number associated with the laser wavelength, and $T=L/c$ is the one-way propagation time inside the cavity of length $L$. $E_\mathrm{in}(t)$ denotes the input optical field, while $E(t)$ is the intracavity electric field. The integer $n$ labels the successive cavity round trips and $N$ is the total number of round trips retained in the calculation. Eventually, the quantity $S_\mathrm{n}(t)$ represents the accumulated optical path over the previous $n$ round trips and accounts for the mirror motion and cavity length variations,
\begin{equation}
    S_n(t) = \sum_{p=0}^{n-1} d(t - 2pT) \,,
\end{equation}
specifically, $d(t) = L +(x_b(t-T)-x_a(t))$ where $x_b(t)$ and $x_a(t)$ are the end and input mirror positions.
Eq.~\ref{eq:1.51} can be used to obtain the optical power $P(t)=|E(t)|^2$ and the \ac{PDH} error signal~\cite{Black2001}, designed for optical cavity control at resonance. In the approximation where the modulation sidebands are fully reflected by the cavity and act as a stable reference field, the demodulated \ac{PDH} signal can be expressed as:
\begin{equation}
    V_\mathrm{PDH}(t) =-\operatorname{Im}\left\{e^{\imath \gamma} E_\mathrm{in}(t)^* E(t)\right\} \,,
    \label{Vpdh}
\end{equation}
where $V_\mathrm{PDH}(t)$ is the \ac{PDH} error signal, and $\gamma$ is the demodulation phase. $E_\mathrm{in}(t)$ is the complex conjugate of the input optical field, while $E(t)$ is the intracavity electric field.

These signals are the optical observables of the cavity state and become strongly non-linear under specific conditions, thereby making standard control technique for locking ineffective. In particular, ring down effect occurs when the mirror velocity becomes sufficiently large. Thresholds to this speed, referred to as  critical velocities, can be defined~\cite{Barsotti2006,swinkelsLockAcquisitionAdvanced2012}. In the present treatment, we consider, as a limit to define the non-linear regime, only the threshold obtained by equating the resonance crossing time to the cavity storage time, which represents the average time a photon remains stored in the cavity. This condition corresponds to the critical velocity defined as follows:
\begin{equation}
   v_\mathrm{cr} \approx \frac{\lambda \pi c}{4L\mathcal{F}^2} \,,
\end{equation}
where, $\lambda$ is the laser wavelength, $c$ is the speed of light, $L$ is the cavity length, and $\mathcal{F}$ is the cavity finesse.

The simulator exploits the structure of Eq.~\ref{eq:1.51} to enable the selection of a sampling frequency over a wide range, as well as the independent motion of both the input and end mirrors. The input mirror displacement is modelled as a phase shift of the input laser field, as adopted in the algorithm.

Because the cavity field equation is recursive, each new sample depends only on an internal simulator state rather than on a full re-evaluation of the entire optical history. The cavity memory is therefore represented by a finite buffer of past input fields together with a single stored values of fields and mirrors position. This mechanism supports incremental, state-space-like updates instead of repeated global re-integration. Geometric electric field decay and phase factors are precomputed once during initialization and reused at every step, so the per-step computational load is dominated by multiply-accumulate operations.

The simulator allows the user to select the sampling frequency, but its final value is approximated according to an internal optimization rule designed to preserve consistency with the cavity round-trip time. 
We will refer to the sampling frequency desired by the user as $f_{\mathrm{calc}}^{\mathrm{desired}}$. The effective sampling frequency $f_{\mathrm{calc}}$ is computed through a rule-based decision tree that compares $f_{\mathrm{calc}}^{\mathrm{desired}}$ with $f_{\mathrm{2T}}=1/2T$, the round trip frequency. The main algorithm pseudocode is shown in Alg.~\ref{alg:strategy_selection}. Alg.~\ref{alg:number_of_subhistories} and Alg.~\ref{alg:round_for_inverse_curve} clarify rounding strategies.

\begin{algorithm}
\caption{number\_of\_subhistories($\eta_{2T}$)}
\label{alg:number_of_subhistories}
\begin{algorithmic}[1]
\Require $\eta_{2T} > 0$
\If{$\eta_{2T} < 1$}
\State \Return $\mathrm{round}(1/\eta_{2T})$
\Else
\State \Return $1$
\EndIf
\end{algorithmic}
\end{algorithm}

\begin{algorithm}
\caption{round\_for\_inverse\_curve($\eta_{2T}$)}
\label{alg:round_for_inverse_curve}
\begin{algorithmic}[1]
\Require $\eta_{2T} > 0$
\State $k_0 \gets \lfloor \eta_{2T} \rfloor$
\State $b \gets \dfrac{2k_0(k_0+1)}{2k_0+1}$
\If{$\eta_{2T} < b$}
\State $k \gets k_0$
\Else
\State $k \gets k_0 + 1$
\EndIf
\end{algorithmic}
\end{algorithm}

In particular, Alg.~\ref{alg:round_for_inverse_curve} implements a custom quantization rule that maps the continuous variable $\eta_{2T}$ to an integer order $k$ used by the simulator. It computes $k_0=\lfloor \eta_{2T}\rfloor$, evaluates the switching threshold
$$
b=\frac{2k_0(k_0+1)}{2k_0+1},
$$
and returns $k=k_0$ if $\eta_{2T}<b$, otherwise $k=k_0+1$. Alg.~\ref{alg:number_of_subhistories} and Alg.~\ref{alg:round_for_inverse_curve} are both used to round the dimensionless round-trip normalized time-step defined as $\eta_{2T} = (f_\mathrm{calc}^\mathrm{desired} 2T)^{-1}$.

Within the full algorithm, this function is the core discretization step. Its role is to choose the integer round-trip count $N$ that best approximates the requested sampling condition, thereby controlling the trade-off between physical fidelity and numerical feasibility of the final computed sampling frequency $f_{\mathrm{calc}}=f_{2T}/N$. We will come back to this function further in the description.

Alg.~\ref{alg:strategy_selection} checks if $f_{\mathrm{calc}}^{\mathrm{desired}}>f_{\mathrm{2T}}$, meaning the requested time-step is shorter than a photon round-trip time. In this case, the algorithm interleaves multiple sub-histories, each shifted in phase to increase the effective sampling, and adjusts $f_\mathrm{calc}$ to an integer multiple of the round-trip frequency.

For $f_{\mathrm{calc}}^{\mathrm{desired}}<f_{\mathrm{2T}}$, we distinguish between two regimes. The first case occurs when the desired sampling frequency becomes too low compared to the cavity storage time; the corresponding integration window may exceed the effective memory of the cavity. In this case, the number of round-trips included in the summation is capped to a maximum value $N_{\max}=5N_{\mathrm{eff}}$, where the effective photon lifetime expressed in round-trips is defined as
\[
N_{\mathrm{eff}} = \frac{1}{\left|\ln(r_a r_b)\right|} \,.
\]
The desired sampling frequency is then preserved, while the integration window is allowed to contain a fractional contribution of the last round-trip, ensuring that negligible long-past contributions are not propagated. This strategy guarantees temporal consistency while balancing numerical efficiency against the physical memory of the cavity.

The second case corresponds to intermediate frequencies, where an integer number of round-trips is selected using an inverse-curve rounding rule that finds optimal $N$, which minimizes the error with respect to the desired sampling frequency. In this case, the effective sampling frequency is set to \(f_{\mathrm{calc}}=f_{2T}/N\).

\begin{algorithm}
\caption{sampling\_frequency\_selection($f_{\mathrm{calc}}^{\mathrm{desired}}, f_{2T}, N_{\mathrm{eff}}$)}
\label{alg:strategy_selection}
\begin{algorithmic}[1]
\State $N \gets 1$
\State $n_{\mathrm{sub}} \gets 1$
\State $\mathrm{partial\_Theta} \gets \mathrm{False}$
\State $\_N_{\mathrm{eff\_factor}} \gets 5$ \Comment{field amplitude has decayed to $\approx 0.67\%$ of initial value}
\State $\eta_{2T} \gets f_{2T} / f_{\mathrm{calc}}^{\mathrm{desired}}$ \Comment{dimensionless round-trip normalized time-step}
\If{$f_{\mathrm{calc}}^{\mathrm{desired}} > f_{2T}$}
\State $n_{\mathrm{sub}} \gets \mathrm{number\_of\_subhistories}(\eta_{2T})$
\State $f_{\mathrm{calc}} \gets f_{2T}\cdot n_{\mathrm{sub}}$
\Else
\State $N_{\max} \gets \_N_{\mathrm{eff\_factor}}\cdot N_{\mathrm{eff}}$
\If{$f_{\mathrm{calc}}^{\mathrm{desired}} < \dfrac{f_{2T}}{N_{max}}$}
\State $N \gets N_{\max}$
\State $f_{\mathrm{calc}} \gets f_{\mathrm{calc}}^{\mathrm{desired}}$
\State $\mathrm{partial\_Theta} \gets \mathrm{True}$
\Else
\State $N \gets \mathrm{round\_for\_inverse\_curve}(\eta_{2T})[0]$
\State $f_{\mathrm{calc}} \gets f_{2T}/N$
\EndIf
\EndIf
\State $\Theta \gets 1/f_{\mathrm{calc}}$
\State \Return $(f_{\mathrm{calc}},\ \Theta,\ N,\ \mathrm{partial\_Theta})$
\end{algorithmic}
\end{algorithm}

\begin{figure}[ht]
    \centering
    \includegraphics[width=0.6\linewidth]{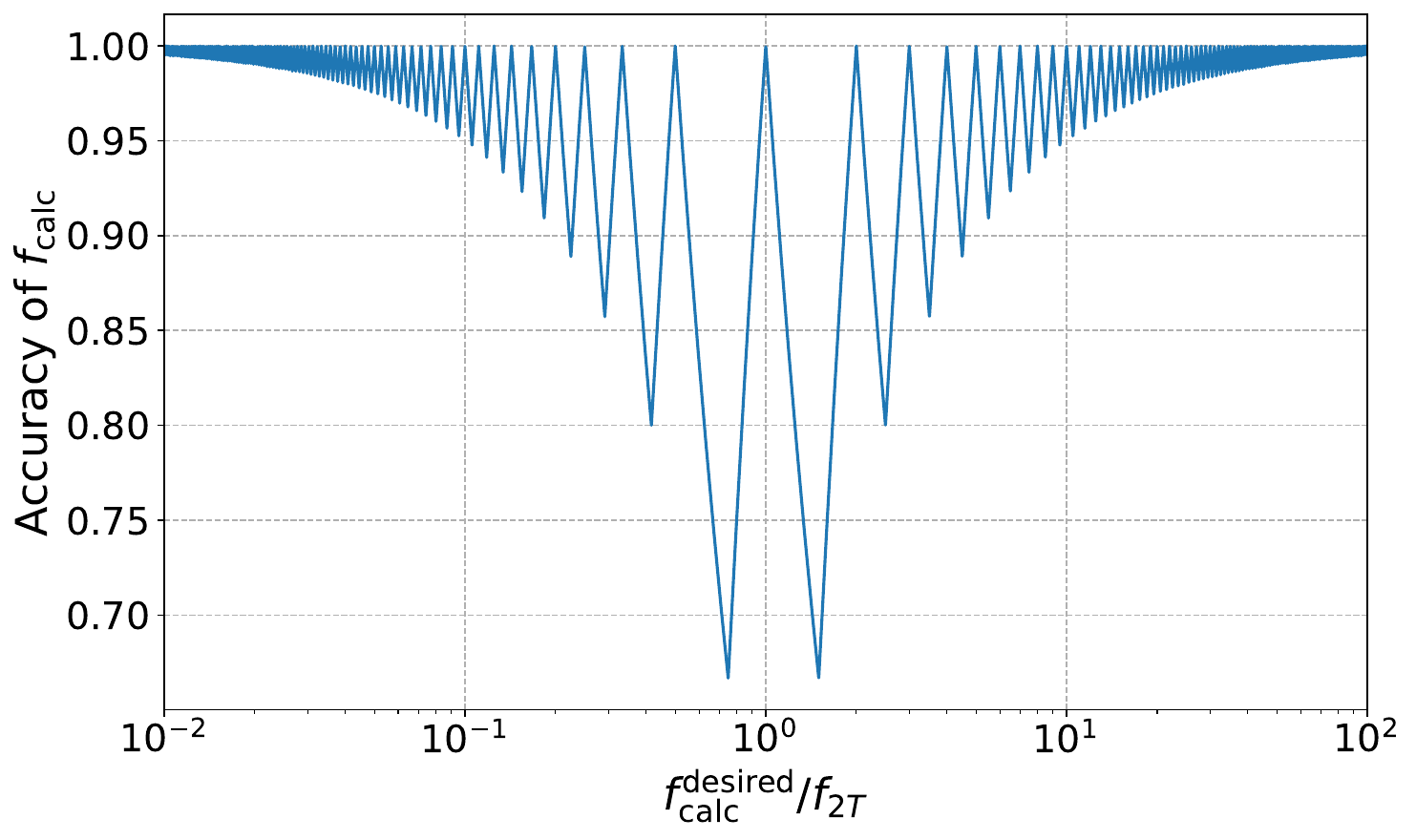}
    \caption{Accuracy of the effective calculation frequency $f_{\mathrm{calc}}$ with respect to the user-defined frequency $f_{\mathrm{calc}}^{\mathrm{desired}}$, as a function of the normalized ratio $f_{\mathrm{calc}}^{\mathrm{desired}}/f_{2T}$.}
    \label{fig:f_calc_accuracy}
\end{figure}

To quantify the impact of the internal frequency-selection strategy, we evaluate the accuracy of the effective sampling frequency with respect to the user-requested value. The result is shown in Fig.~\ref{fig:f_calc_accuracy}. The plot shows the relative accuracy
\[
A_f = 1-\frac{\left|f_{\mathrm{calc}}-f_{\mathrm{calc}}^{\mathrm{desired}}\right|}{f_{\mathrm{calc}}^{\mathrm{desired}}}\,,
\]
highlighting the minimal achievable deviation introduced by the approximation to values compatible with the cavity round-trip time.

The simulator workflow can be summarized in a few simple stages. 
First, the cavity is defined by specifying its physical parameters: mirror reflectivities, transmissivities, and cavity length, together with the desired sampling frequency $f_{\mathrm{calc}}^{\mathrm{desired}}$. 
During the initialization phase, the internal frequency-selection rule determines the effective sampling frequency $f_{\mathrm{calc}}$, the corresponding time step $\Theta$, and the number of round-trips included in each integration step. 
The effective sampling frequency is then returned to the user and must be used as the temporal grid of any external simulation environment. 
The simulation proceeds iteratively, updating the intracavity field at each step through the recursive field equation.

\section{Validation and benchmarks}
\label{sec:Validation_and_benchmarks}
The simulator performance is evaluated through comparison with experimental data acquired from the north arm cavity of the Virgo interferometer~\cite{virgo_optical_params_2007, std_Virgo_ref_Acernese_2023} during mode matching measurement using the external excitation method, which consists of applying corrections to excite mechanically the end mirror in the longitudinal degree of freedom and analysing the second order mode during free swinging of the test mass. This mechanical excitation produces non-adiabatic crossings of the cavity resonances, exposing ringing effect for different crossing velocities. These represents optimal conditions for validating the simulation framework.

The selected dataset corresponds to a \SI{10}{\second} time window of the DC transmission power signal recorded at the cavity auxiliary B7 photodetector. Together with the transmission power, \ac{PDH} error signal data from the B4 photodetector in reflection has been collected to compare with simulated data obtained from Eq.~\ref{Vpdh}.

\begin{figure}[ht]
    \centering
    \includegraphics[width=0.8\linewidth]{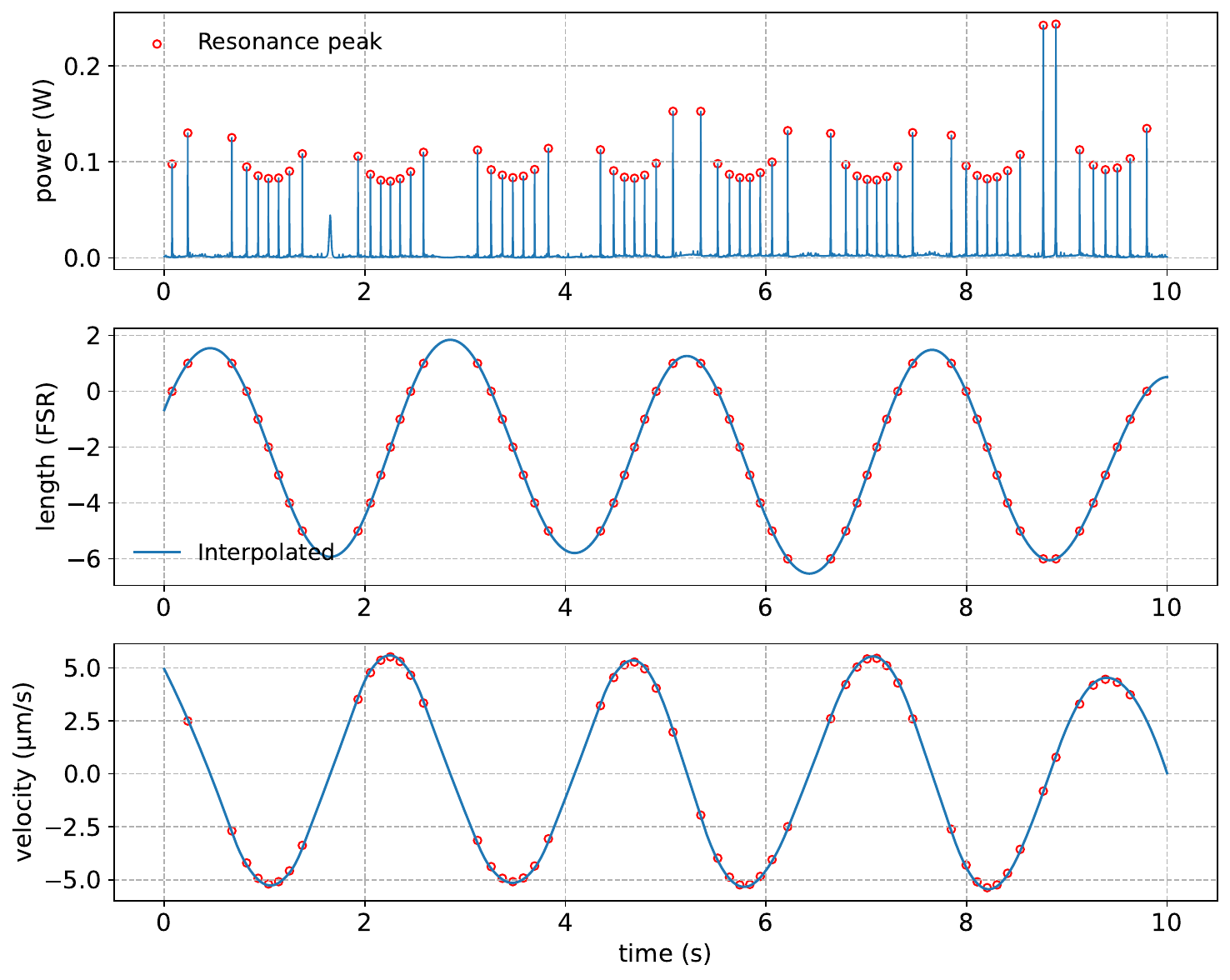}
    \caption{Cavity length and velocity reconstruction during mirror oscillation following a mechanical excitation. All panels share the same horizontal axis, representing time. The upper panel shows the recorded transmission peaks in the experimental data; peaks used in the analysis are indicated by circles. The middle panel depicts the cavity length in \ac{FSR} units and the corresponding polynomial fit. The bottom panel shows the cavity mirror velocity derived from the cavity length. The first point and the last point of this pane has been cut due to edge effect caused by numerical derivative calculation.}
    \label{fig:speed_calc}
\end{figure}

The experimental data are shown in Fig.~\ref{fig:speed_calc} where it is also depicted the result of its processing. First, the first sample timestamp was set to zero. Then the transmission was detrended by subtracting its median value. Finally, resonance peak identification was performed with an algorithm based on a prominence-based peak detection: only peaks above $99^\mathrm{th}$ percentile of the detrended signal were retained. Each detected peak corresponds to a resonance condition $2L(t_n) = N_n \lambda$, with $\lambda = \SI{1064}{\nano\metre}$. Successive resonances therefore correspond to cavity length variations of $\Delta L = {\lambda}/2$.

An integer resonance index $n$ was assigned to each detected peak to reconstruct the cavity-length variation. The direction of the first crossing was determined from the sign of the local derivative of the \ac{PDH} signal, after which the index was increased or decreased by one at each subsequent resonance. Turning points were identified based on a threshold for the transmission, which will be higher when the mirror velocity becomes small before changing direction. At these points, the sign of the index increment was reversed. The absolute cavity length for each $n$ was reconstructed and expressed in terms of unity of \ac{FSR}.
The discrete length samples were interpolated using a cubic spline to obtain a continuous estimate $L(t)$ over the oscillation window. A cubic-spline interpolation provides a smooth reconstruction of the resonance positions, but it is not equivalent to a sinusoidal model. Unlike a sinusoidal fit, the spline does not impose a global physical model on the mirror motion and may therefore introduce non-physical distortions in the derived velocity. For this reason, spline-based reconstruction is useful as a flexible interpolation tool, but can affect the velocity estimation.

The instantaneous mirror velocity was computed from the time derivative of the reconstructed cavity length $v(t) = \mathrm{d}L(t)/\mathrm{dt}$.

The computation of the simulated data was performed by initializing the simulation with cavity parameters consistent with arm cavity of the Virgo interferometer~\cite{capocasaOpticalNoiseStudies}: mirror reflectivities $R_a = 0.986$, $R_b = 0.99999$, input transmission $T_a = 0.01377$, and nominal length $L = \SI{3000}{\metre}$. The input laser field was set to a mean value of $|E_{\mathrm{in}}|= \num{12,7}$, obtained by the least squares method fit described below. The calculation frequency was matched to the sampling frequency of the raw data which is \SI{10}{\kilo\hertz}, ensuring temporal resolution well below the resonance crossing time. It is important to specify that in this case the actual calculation frequency approximated by the simulation strategy was $\approx \SI{9,99}{\kilo\hertz}$, with an approximation accuracy of $\SI{99,9}{\percent}$. With this values only one history of the optical field evolution was evaluated and the regime was the aggregated round-trips one, considering the full Eq.~\ref{eq:1.51}.

The reconstructed velocity was used as input to the simulation, defining a time-dependent cavity length variation. The simulated cavity power was obtained from the intra-cavity electric field through the output mirror transmissivity.

\begin{figure}[ht]
    \centering
    \includegraphics[width=0.8\linewidth]{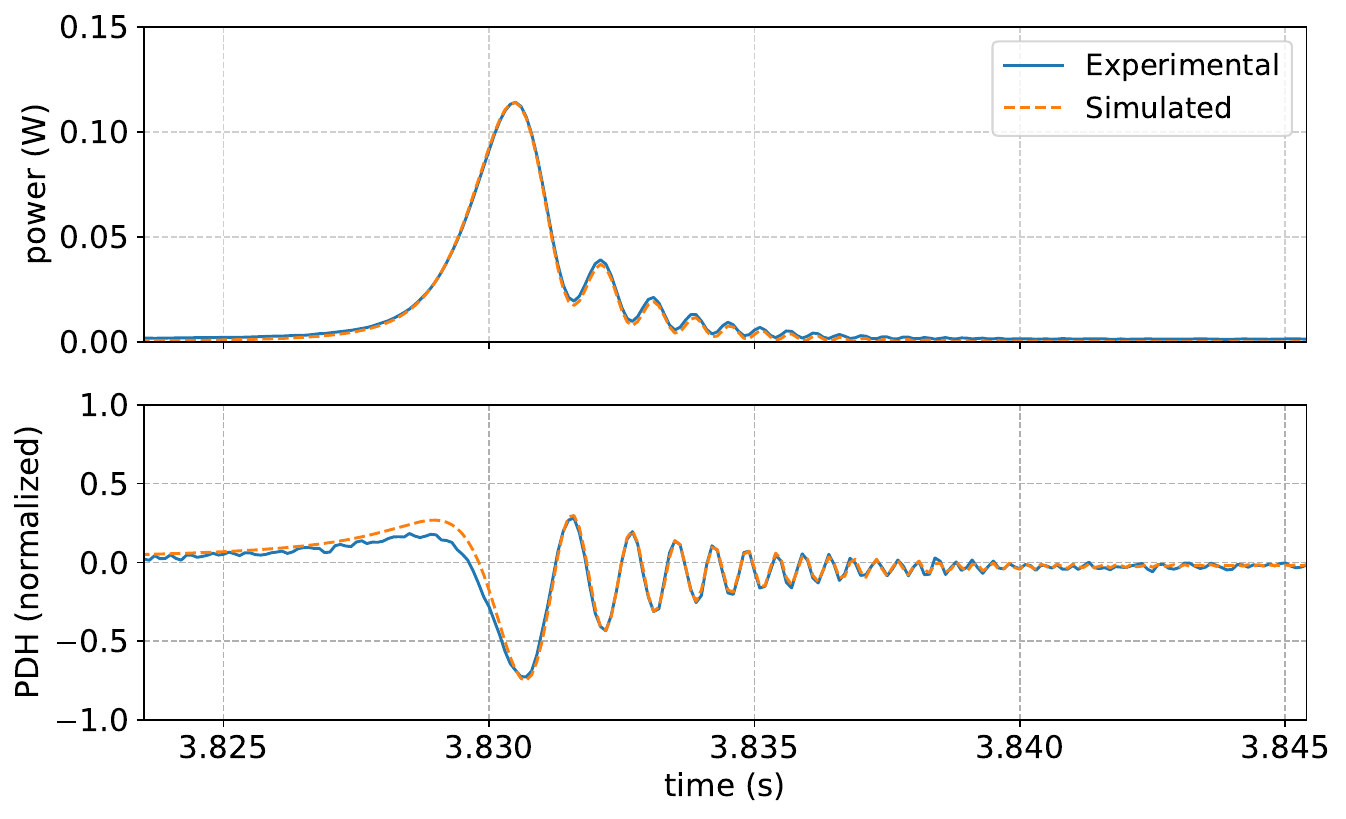}
    \caption{Comparison between experimental data and simulation for a single cavity resonance crossing. The top panel shows the experimental and simulated DC transmission, while the bottom panel displays the corresponding normalized \ac{PDH} error signal. The simulation is driven by the reconstructed mirror motion and reproduces both the amplitude and temporal structure of the resonance, using a critical velocity scaling factor of $7.6$, which corresponds to $\SI{4,9}{\micro\metre\per\second}$}
    \label{fig:single_res}
\end{figure}

For detailed analysis, one resonance has been chosen, corresponding to $\approx \SI{4.9}{\micro\metre\per\second}$, \textit{i.e.} a factor of $\approx 7.6$ critical velocities. The compared data are showed in Fig.~\ref{fig:single_res} \\ 
In addition to the DC transmission, the simulated and reconstructed \ac{PDH} error signals exhibit consistent zero-crossing slope and phase structure, confirming that the simulator correctly captures the dynamics induced by high-velocity mirror motion.

In the appendix we present an additional plot for validation of the simulator which consists of a comparison of the whole experimental data time window ($\SI{10}{\second}$) with some observations about integration errors.

In order to find the laser input power, Eq.~\ref{eq:1.51} can be approximated by a sum of adiabatic contribution plus a Doppler term:
\begin{equation}
    E(t) \approx D_0 \exp{\left(-\dfrac{t}{\tau} - \imath\dfrac{kv}{2T} t^2\right)} + \frac{t_a E_{\mathrm{in}}}{1-r_a r_b e^{-2\imath kvt}} \,,
    \label{eq:2.69}
\end{equation}
where, $D_0$ is the complex amplitude of the transient contribution, $\tau$ is the cavity storage time, $v$ is the mirror velocity, and $E_\mathrm{in}$ is the constant incident field. The prefactor $D_0$ is given by:
\begin{equation}
    D_0 = t_a E_{\mathrm{in}} \sqrt{\frac{\imath \pi}{2kvT}} \exp{\left(\dfrac{\imath T}{2kv\tau^2}\right)} \,.
\end{equation}

This approximation holds in the length-scan transient regime for $v \ge v_{cr}$. Its derivation assumes a continuum limit of the discrete round-trip dynamics $n \gg 1$ and timescales $t \gg T$. The immediate resonance region is neglected~\cite{Rakhmanov2000}; therefore, the least-squares fit is performed only on the tail of the ring-down.

In addition to the physical validation against experimental data, the computational performance of the simulator has been systematically assessed through a dedicated benchmark study. The benchmark on almost 400 cases compares the pure Python and Numba-accelerated backends on a test set of 99 cavity configurations spanning different lengths, reflectivities, and calculation frequencies. In all cases, the execution time scales approximately linearly with the number of round trips $N$, reflecting the dominant cost of the summation over stored field history. The pure Python implementation shows a clear increase in runtime with growing $N$, reaching several hundred microseconds for high-finesse, short cavities. In contrast, the Numba backend exhibits a much weaker dependence on $N$, reaching \SI{16}{\micro\second} for the most demanding cavity tested, with $N=300$.

As a result, the speed-up factor increases systematically with $N$, ranging from \num{4.67} for $N\approx1$ up to $17.61$ for $N=300$, demonstrating that \ac{JIT} compilation is necessary for efficient simulations in the high-finesse regime. Detailed results are shown in Tab.~\ref{tab_benchmark}. 

\section{Relevance to optics and photonics}
\label{sec:Optical and Photonic relevance}

Beyond \ac{FP} cavities, the time-domain simulator developed in this work may be of relevance to a broader class and physical systems characterized by energy storage, ringing effects, and finite propagation time, all of which exhibit memory-driven dynamics analogous to those of optical cavities. 
Within this framework, the same delay-based formulation naturally extends to integrated optical resonators, such as ring resonators~\cite{Little_1997}, which can be treated as a unidirectional cavities where the intra-cavity field is recursively re-injected after each round-trip with attenuation and phase accumulation.
Ring resonators constitute key building blocks of integrated photonic, including frequency-comb micro-resonators, Kerr non-linear devices, optical buffers, all-optical switching architectures, and quantum photonic circuits.

As an additional validation of the generality of our time-domain approach, we reproduce the characteristic ring-down dynamics obtained in~\cite{Yong_Lee_1999}, namely the transition between pulsed and smooth exponential decay regimes as a function of pulse duration and detuning. Fig.~\ref{fig:Lee-image} in the Appendix~\ref{appendix} presents simulated ring-down signals for different normalized pulse durations and detunings, showing qualitative agreement with the temporal feature described in~\cite{Yong_Lee_1999}, including the emergence of transient peaks and intensity modulation arising from the partial temporal overlap of successive photon round trips.

Another promising application of the simulator lies in the optimization of locking schemes for extremely high-finesse cavities ($\mathcal{F} \sim \num{10000}$). In this regime, resolving the temporal evolution of the intra-cavity field is crucial to understand transient dynamics and control performance. Machine learning techniques, particularly \ac{RL}, can exploit this information to identify optimal control strategies, avoiding direct interaction with experimental hardware, which can be both time-consuming and potentially harmful~\cite{Svizzeretto_2026, bawaj2026_arxiv}.

Finally, such simulations can support the design and optimization of gravitational-wave interferometers such as Virgo and Einstein Telescope, improving their design and further upgrades~\cite{Dorigo_2023,Punturo_2010,Hild_2011}. In particular, in a situation where phenomena like transient dynamics, mirror motion, radiation pressure effects,play a key role in maintaining stable operation.

\section{Conclusions and applications}
\label{sec:Conclusions_and_Applications}

In this work we presented the logic and the validation of a fast time domain optical simulator, capable of reproducing the dynamical evolution of the intra-cavity field in optical resonators. The simulator allows the user to flexibly define cavity parameters, boundary conditions, and mirror motion.

Computational efficiency is achieved through a simple yet effective simulation strategy, in which the user-defined sampling frequency is approximated to the nearest value compatible with internal optimization constraints. This adjustment ensures consistency between the discrete time step and the cavity round-trip time, preserving causal propagation while avoiding unnecessary evaluations of redundant round-trip histories. Efficiency is further improved thanks to the use of \ac{JIT} compilation.

The resulting framework is designed to provide a computationally efficient and physically consistent simulation tool, spanning applications from large-scale interferometric systems to integrated photonic devices.
Computational efficiency is particularly important for predictive modelling, but it becomes even more critical when considering real-time control development and design of \ac{RL}-based lock acquisition protocols. Lock acquisition in suspended optical cavities is intrinsically a non-stationary control problem, in which the cavity state is observed through signals that deviate from the linear regime assumed by classical controllers. 

The feature of a step-by-step evolution of the intracavity field and its associated observables (DC transmission, \ac{PDH} error signal), under arbitrary time-dependent mirror motion and input field modulation, naturally matches the interaction paradigm required by modern \ac{RL} algorithms, where an agent observes the system state at discrete time steps, applies an action (e.g., actuation on mirror position or laser frequency), and receives a reward based on the effectiveness of its actions in reaching the desired goal.

The development of framework empowered with this simulator enables large-scale training campaigns, in which thousands of interaction steps are required to learn robust policies. Moreover, addressing one of the biggest challenge in \ac{RL} applications, the so called sim-to-real transfer, requires high-fidelity simulation of the physical system. Such fidelity is mandatory to train \ac{RL} agents under realistic dynamics, thereby ensuring a safe and more reliable deployment on the real optical setups.

\section*{Acknowledgements}

The authors acknowledge the use of technical data and interferometer parameters from the Virgo Collaboration for validation and comparison purposes. The authors thank the collaboration members involved in detector commissioning and characterization activities for making these data available.

\paragraph{Funding information}
Research is supported by Italian Ministry of University and Research (MUR) through the program "Dipartimenti di Eccellenza 2023-2027" (Grant SUPER-C), Italy.
\\~
This work was partially supported by the European Union’s Horizon Europe programme under the Marie Sk\l odowska-Curie Actions Staff Exchanges project GRAVITY (G.A. 101236384).

\printbibliography

\newpage
\appendix
\section{Auxiliary materials for validation}
\label{appendix}

\begin{figure}[htbp]
    \centering
    \includegraphics[width=0.8\linewidth]{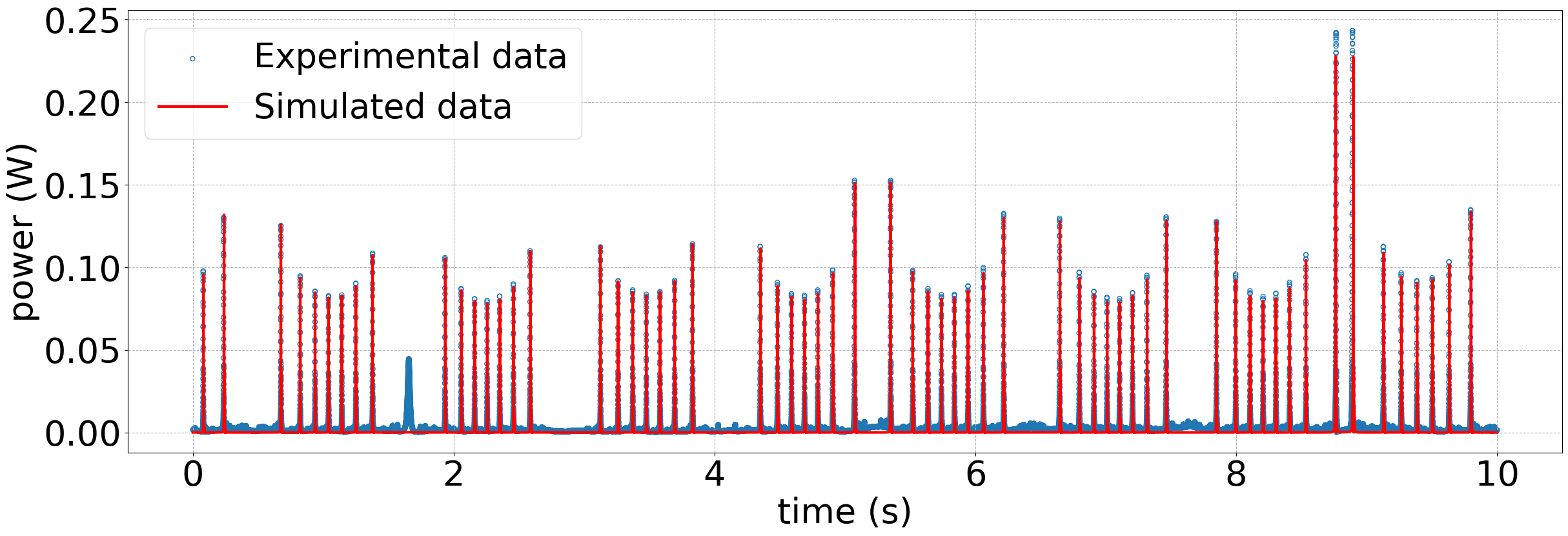}
    \caption{Comparison between the measured DC transmission of the North arm cavity following a mechanical excitation of the suspended mirror and the corresponding simulation. The simulated response reproduces the sequence of resonance crossings induced by the mirror oscillation, using the reconstructed speed evolution as input.}
    \label{fig:kick_comparison}
\end{figure}

Fig.~\ref{fig:kick_comparison} shows the comparison between the measured DC transmission during the mode-matching measurements and the corresponding simulated data for the full $\SI{10}{\second}$ time window. It is important to note that the simulator is not specifically designed for long-time propagations, as the numerical implementations relies on discrete time steps and finite round-trip summations. Consequently, small integration errors, together with finite numerical precisions, accumulate over time and may lead to a gradual degradation of accuracy. 

Despite these limitations, the agreement is remarkably good: the simulated response accurately reproduces both the sequence and relative amplitudes of the resonance crossings.

Part of the residual discrepancy also arises from the resonance identification procedure described in Sec.~\ref{sec:Validation_and_benchmarks}. Resonance positions were estimated by selecting transmission peaks above the \SI{99}{\percent} percentile of the signal and assigning them as the exact resonance crossing times. This assumption becomes less accurate as the mirror velocity increases. In the non-adiabatic regime, the first power peak does not, in general, temporarily coincide with the exact resonance position. This temporal offset propagates into the reconstructed cavity length and, consequently, into the velocity estimation.

Therefore, the observed deviations between simulation and experiment should be interpreted as the combined effect of numerical integration errors and systematic uncertainties in the resonance localization procedure.

\begin{figure}[htbp]
    \centering
    \includegraphics[width=0.9\linewidth]{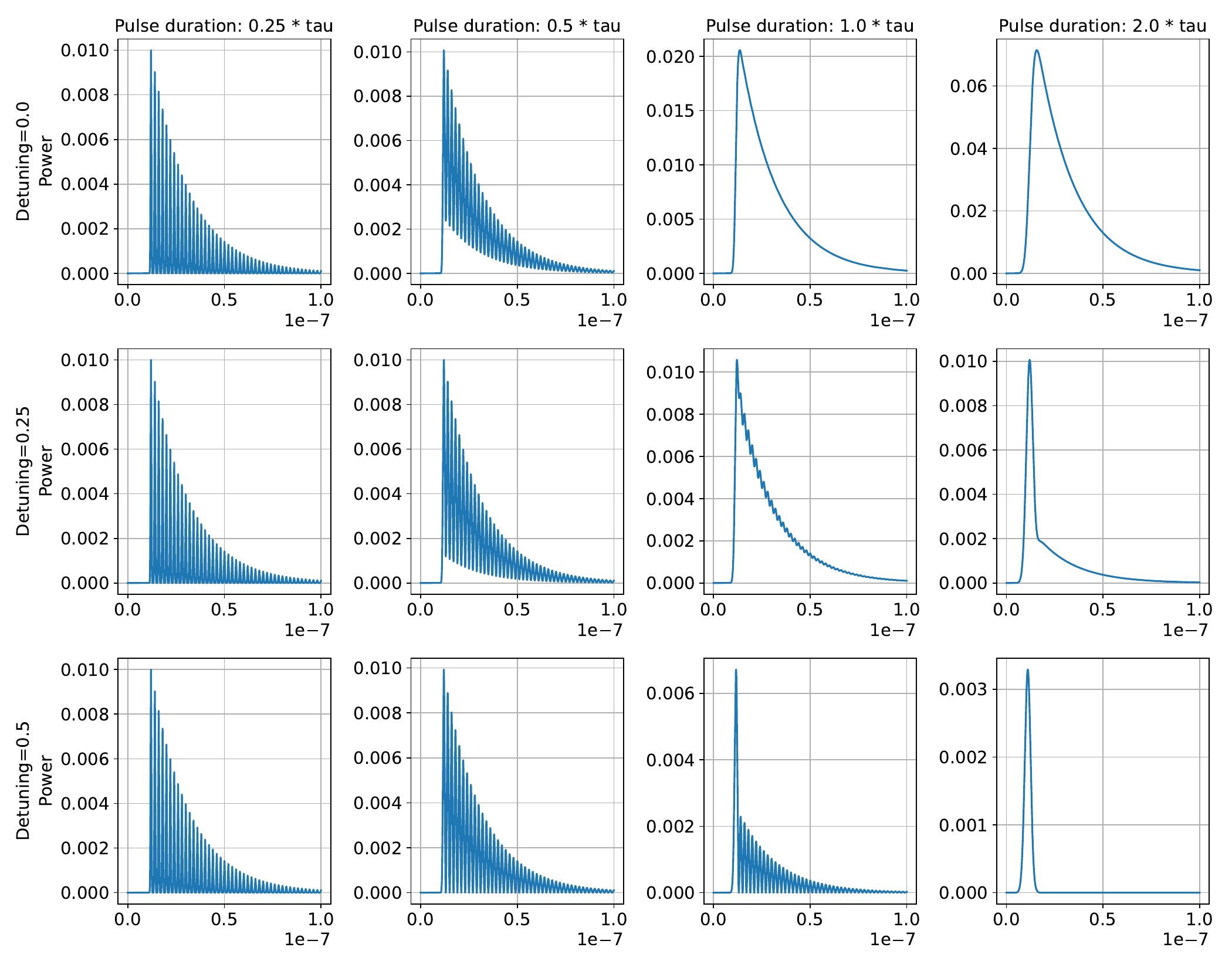}
    \caption{Reproduction of Fig.~2 from~\cite{Yong_Lee_1999}: simulated cavity ring-down signals under pulsed laser excitation for different ratios of pulse duration to cavity round-trip time. The simulator reproduces all reported regimes: well-separated replicas, partial temporal overlap with interference modulation, and fully overlapped excitation leading to a smooth exponential decay. In this reproduction, a sampling frequency of \SI{2}{\tera\hertz} was used, and \num{2e5} simulation points were generated for each panel.}
    \label{fig:Lee-image}
\end{figure}

 Our simulator reproduces results from~\cite{Yong_Lee_1999} using the same physical parameters and without ad hoc fitting. This agreement validates the time-domain modelling of round-trip summation, phase accumulation, and cavity decay, and demonstrates that the published result emerges as a limiting case of the more general simulation framework.

\newpage
\begin{longtable}{|l|l|l|l|l|l|l|l|l|}
    \caption{Table with benchmark results on the full set of test cavities}
    \label{tab_benchmark}\\
    \hline
        \hline
        N & Count & \makecell[l]{Speedup factor\\(mean$\pm$std)} & \makecell[l]{Memory ratio\\(mean)} & \makecell[l]{Pure time\\(\si{\micro\second})} & \makecell[l]{Numba time\\(\si{\micro\second})} & \makecell[l]{Pure mem\\(\si{\kilo\byte})} & \makecell[l]{Numba mem\\(\si{\kilo\byte})} \\ \hline
        \hline
        1 & 140 & $4.67 \pm 0.07$ & 6.48 & 26.18 & 5.61 & 6.98 & 1.08 \\ \hline
        2 & 26 & $5.09 \pm 0.26$ & 6.60 & 29.03 & 5.74 & 7.27 & 1.10 \\ \hline
        3 & 32 & $5.29 \pm 0.11$ & 6.49 & 29.86 & 5.65 & 7.30 & 1.12 \\ \hline
        4 & 6 & $5.38 \pm 0.07$ & 6.37 & 30.64 & 5.70 & 7.32 & 1.15 \\ \hline
        5 & 14 & $5.60 \pm 0.35$ & 6.27 & 32.10 & 5.74 & 7.34 & 1.17 \\ \hline
        6 & 20 & $5.64 \pm 0.07$ & 6.18 & 32.45 & 5.76 & 7.39 & 1.20 \\ \hline
        7 & 11 & $5.74 \pm 0.06$ & 6.06 & 33.15 & 5.78 & 7.39 & 1.22 \\ \hline
        8 & 3 & $5.76 \pm 0.09$ & 5.97 & 34.08 & 5.91 & 7.41 & 1.24 \\ \hline
        9 & 2 & $5.94 \pm 0.03$ & 5.88 & 34.52 & 5.82 & 7.44 & 1.27 \\ \hline
        10 & 1 & $6.11$ & 5.79 & 35.98 & 5.89 & 7.46 & 1.29 \\ \hline
        11 & 12 & $6.16 \pm 0.10$ & 5.70 & 36.38 & 5.91 & 7.48 & 1.31 \\ \hline
        12 & 3 & $6.37 \pm 0.03$ & 5.62 & 37.55 & 5.89 & 7.51 & 1.34 \\ \hline
        13 & 7 & $6.46 \pm 0.15$ & 5.54 & 38.23 & 5.92 & 7.53 & 1.36 \\ \hline
        14 & 6 & $6.56 \pm 0.06$ & 5.46 & 39.04 & 5.95 & 7.55 & 1.38 \\ \hline
        16 & 2 & $6.79 \pm 0.06$ & 5.32 & 40.89 & 6.03 & 7.60 & 1.43 \\ \hline
        17 & 3 & $6.87 \pm 0.10$ & 5.25 & 41.32 & 6.01 & 7.62 & 1.45 \\ \hline
        18 & 6 & $6.98 \pm 0.07$ & 5.18 & 42.52 & 6.09 & 7.65 & 1.48 \\ \hline
        19 & 1 & $7.06$ & 5.11 & 43.33 & 6.14 & 7.67 & 1.50 \\ \hline
        22 & 1 & $7.47$ & 4.93 & 45.95 & 6.15 & 7.74 & 1.57 \\ \hline
        23 & 2 & $7.32 \pm 0.01$ & 4.87 & 46.21 & 6.31 & 7.77 & 1.59 \\ \hline
        24 & 1 & $7.56$ & 4.82 & 80.24 & 10.61 & 7.79 & 1.62 \\ \hline
        25 & 2 & $7.68 \pm 0.11$ & 4.76 & 48.72 & 6.34 & 7.81 & 1.64 \\ \hline
        26 & 1 & $7.87$ & 4.71 & 49.27 & 6.26 & 7.84 & 1.66 \\ \hline
        27 & 4 & $7.74 \pm 0.07$ & 4.66 & 49.29 & 6.37 & 7.86 & 1.69 \\ \hline
        28 & 9 & $7.89 \pm 0.05$ & 4.61 & 51.05 & 6.47 & 7.88 & 1.71 \\ \hline
        30 & 12 & $8.09 \pm 0.07$ & 4.51 & 52.44 & 6.48 & 7.93 & 1.76 \\ \hline
        31 & 8 & $8.17 \pm 0.11$ & 4.46 & 53.25 & 6.52 & 7.95 & 1.78 \\ \hline
        38 & 1 & $8.77$ & 4.17 & 58.81 & 6.71 & 8.12 & 1.95 \\ \hline
        46 & 1 & $9.59$ & 3.89 & 65.85 & 6.87 & 8.30 & 2.13 \\ \hline
        47 & 1 & $9.55$ & 3.86 & 65.93 & 6.90 & 8.33 & 2.16 \\ \hline
        50 & 1 & $9.76$ & 3.77 & 69.42 & 7.11 & 8.40 & 2.23 \\ \hline
        53 & 1 & $10.05$ & 3.69 & 71.55 & 7.12 & 8.47 & 2.30 \\ \hline
        55 & 5 & $10.21 \pm 0.09$ & 3.63 & 74.01 & 7.25 & 8.52 & 2.34 \\ \hline
        57 & 1 & $10.34$ & 3.58 & 74.04 & 7.16 & 8.56 & 2.39 \\ \hline
        60 & 12 & $10.35 \pm 0.11$ & 3.51 & 77.21 & 7.46 & 8.63 & 2.46 \\ \hline
        62 & 2 & $10.77 \pm 0.02$ & 3.46 & 79.23 & 7.36 & 8.68 & 2.51 \\ \hline
        63 & 5 & $10.73 \pm 0.14$ & 3.44 & 79.96 & 7.45 & 8.70 & 2.53 \\ \hline
        78 & 1 & $11.73$ & 3.14 & 93.75 & 7.99 & 9.05 & 2.88 \\ \hline
        96 & 1 & $12.57$ & 2.88 & 111.52 & 8.87 & 9.52 & 3.30 \\ \hline
        102 & 1 & $13.09$ & 2.80 & 115.24 & 8.80 & 9.66 & 3.45 \\ \hline
        104 & 1 & $12.94$ & 2.78 & 116.89 & 9.04 & 9.71 & 3.49 \\ \hline
        108 & 1 & $13.36$ & 2.73 & 120.86 & 9.05 & 9.80 & 3.59 \\ \hline
        112 & 1 & $13.68$ & 2.69 & 125.01 & 9.14 & 9.90 & 3.68 \\ \hline
        118 & 1 & $13.72$ & 2.63 & 128.56 & 9.37 & 10.04 & 3.82 \\ \hline
        132 & 1 & $14.27$ & 2.50 & 139.91 & 9.81 & 10.37 & 4.15 \\ \hline
        136 & 1 & $14.50$ & 2.47 & 145.02 & 10.00 & 10.46 & 4.24 \\ \hline
        137 & 1 & $14.41$ & 2.45 & 140.53 & 9.75 & 10.44 & 4.27 \\ \hline
        139 & 1 & $14.81$ & 2.43 & 144.25 & 9.74 & 10.48 & 4.31 \\ \hline
        140 & 4 & $14.78 \pm 0.09$ & 2.42 & 144.96 & 9.81 & 10.51 & 4.34 \\ \hline
        166 & 1 & $15.62$ & 2.26 & 169.71 & 10.86 & 11.16 & 4.95 \\ \hline
        176 & 1 & $15.77$ & 2.20 & 176.53 & 11.19 & 11.40 & 5.18 \\ \hline
        212 & 1 & $16.77$ & 2.03 & 206.86 & 12.34 & 12.24 & 6.02 \\ \hline
        256 & 1 & $17.37$ & 1.88 & 248.35 & 14.29 & 13.27 & 7.05 \\ \hline
        276 & 4 & $17.56 \pm 0.09$ & 1.82 & 262.30 & 14.93 & 13.70 & 7.52 \\ \hline
        300 & 8 & $17.61 \pm 0.10$ & 1.76 & 282.15 & 16.02 & 14.26 & 8.09 \\ \hline
        \hline
\end{longtable}

\end{document}